\documentclass[11pt,a4paper]{article}

\usepackage[T1]{fontenc}
\usepackage[utf8]{inputenc}
\usepackage{mathptmx}
\usepackage{amsmath,amssymb}
\usepackage{graphicx}
\usepackage{textcomp}
\usepackage[margin=2.5cm]{geometry}
\usepackage{caption}
\usepackage{authblk}
\usepackage[numbers,square,sort&compress]{natbib}

\usepackage[colorlinks=true,linkcolor=blue,citecolor=blue,urlcolor=blue]{hyperref}

\hypersetup{
  pdftitle={Metasurface-integrated VCSEL designed for polarization control in optical Ising machines},
  pdfauthor={Wenjie Chen, Zifeng Yuan, Hong-Lin Lin, Luo Qi, Jiaru Chu, Aaron Danner, Yuhang Chen},
  pdfsubject={Vertical-cavity surface-emitting lasers; metasurfaces; polarization control; photonic Ising machines},
  pdfkeywords={metasurface, VCSEL, polarization control, anisotropy, bireflectance, spin-flip model, optical injection locking, Ising machine, photonic computing}
}

\title{\bfseries Metasurface-integrated VCSEL designed for polarization control in optical Ising machines}

\author[1,2,3]{Wenjie Chen}
\author[3]{Zifeng Yuan}
\author[3]{Hong-Lin Lin}
\author[3,*]{Luo Qi}
\author[1,2]{Jiaru Chu}
\author[3]{Aaron Danner}
\author[1,2,*]{Yuhang Chen}
\affil[1]{Department of Precision Machinery and Precision Instrumentation, University of Science and Technology of China, Hefei 230027, China}
\affil[2]{Key Laboratory of Precision Scientific Instrumentation of Anhui Higher Education Institutes, University of Science and Technology of China, Hefei 230027, China}
\affil[3]{Department of Electrical and Computer Engineering, National University of Singapore, Singapore 117583, Singapore}
\affil[*]{\href{mailto:eleql@nus.edu.sg}{eleql@nus.edu.sg} and \href{mailto:chenyh@ustc.edu.cn}{chenyh@ustc.edu.cn}}

\date{}

\begin{document}
\maketitle

\begin{abstract}
\noindent
The orthogonal polarization states of vertical-cavity surface-emitting lasers (VCSELs) can be used to describe candidate solutions to the Ising Hamiltonian, which is useful for solving quadratic unconstrained binary optimization problems. However, the natural anisotropy of VCSELs tends to overly favor one polarization state, which impedes the system from working as desired. In this work, we have designed and fabricated a metasurface, which may lead to a VCSEL with reduced undesired anisotropy. By changing the geometric size of nano-structures in the metasurface, the polarization state of the output light can be altered. Based on the injection-locking theory and spin-flip model, we numerically show that VCSELs with lowered anisotropy are more easily affected by the injection locking needed in Ising systems. Additionally, we numerically study the evolution of a 3-bit VCSEL-based Ising system and verify that the computational accuracy of the photonic Ising machine can be improved to more than twice that of its counterpart with higher anisotropy.
\end{abstract}

\section{Introduction}

Owing to the low power consumption, high computational speed and promise of parallel computing, optical computing systems are particularly suitable for future solving of complex large-scale problems. To date, optical computing has been widely applied in solving combinatorial optimization problems~\cite{Kyriienko2019, Wang2021, Mwamsojo2023, Gao2024}, calculating inverse matrices~\cite{Chen2022inv, Chen2024inv}, performing matrix multiplications~\cite{Lin2018, Yan2019, Zhou2022, Liu2025}, conducting Fourier transformations~\cite{Goodman1978, Zhu2022}, etc. Among these, a hypothetical ability to solve combinational optimization problems with high throughput and low latency is of great importance in fields like communication, molecular analysis, circuit design, etc. More broadly, the demand for such hardware is driven by increasingly compute-intensive machine-learning and computer-vision workloads, ranging from polarization-aware imaging~\cite{lin2025rgb} and 3D scene reconstruction~\cite{Teng2025} to vision-language reasoning~\cite{Du2026} and language-model-driven scientific discovery~\cite{Chen2026hypo, chen2025auto}.

The term Ising model originates from magnetic materials. The Ising model consists of a lattice of spins, where each spin takes a value of either ``$+1$'' or ``$-1$'', representing the binary spin states of spin up and spin down~\cite{Ising1925}. As a physical system of $M$ ($M \in \mathbb{N}$) coupled spins spontaneously evolves towards its ground state with minimum energy, each spin (also known as an Ising bit) will evolve into a certain state eventually. The ground state of the system corresponds to the minimum value of Hamiltonian, described by
\begin{equation}
H = -\sum_{i<j}^{M} J_{i,j}\,\sigma_i \sigma_j + \sum_{i=1}^{M} h_i \sigma_i
\label{eq:ising}
\end{equation}
where $\sigma$ is the spin state, and $\sigma$ has a value of either $+1$ or $-1$. $J_{i,j}$ is the interaction coefficient between spin $i$ and spin $j$, and the Zeeman term $h_i\sigma_i$ represents an external influence on spin $i$.

An Ising machine is a computational system based on the Ising model. Simulating the natural behavior of the physical system, the process of finding a solution to a quadratic unconstrained binary optimization problem is mathematically equivalent to finding the ground state of the Ising model. Therefore, a combinational optimization problem can be solved by an Ising machine via mapping it to the Ising model. Along with circuit-based Ising machines~\cite{Yamaoka2016, Merolla2014, Tsukamoto2017}, there are photonic Ising machines~\cite{Utsunomiya2011, Takata2014, Utsunomiya2015, Cen2022, Inagaki2016a, McMahon2016, Marandi2014, Inagaki2016b, Takesue2016, Babaeian2019, Tezak2020, Okawachi2020, Prabhu2020, Gao2024} as well as Ising machines, which make use of quantum mechanical principles to reach the ground state (which is then termed quantum annealing)~\cite{Santoro2002, Albash2018, Mott2017, Hamerly2019}. For photonic Ising machines, physical parameters with binary values, like polarization, amplitude, phase, can be selected as candidates for the spin state variable in the Ising model. To date, photonic Ising machines have been realized based on multiple systems, including injection-locked lasers~\cite{Utsunomiya2011, Takata2014, Utsunomiya2015}, VCSEL networks with parallel optical feedback~\cite{Lim2024, Zhang2025b, Zhang2025}, optical parameter oscillators~\cite{Cen2022, Inagaki2016a, McMahon2016, Marandi2014, Inagaki2016b, Takesue2016}, multicore fibers~\cite{Babaeian2019}, micro-ring resonators~\cite{Tezak2020, Okawachi2020}, Mach-Zehnder interferometers~\cite{Prabhu2020}, etc.

Having advantages of low cost, low threshold current, high electro-optical conversion efficiency, and ease of integration into 2D arrays, vertical cavity surface emitting lasers (VCSELs) have been widely applied to high-volume and high-speed optical fiber communication systems~\cite{Iga2008, Hadley1995}. Additionally, the polarization of light emitted by VCSELs can exhibit bi-stability, usually referred to two orthogonal directions, which we shall term the ``X-polarization'' and ``Y-polarization''. A lasing state can flip from one to another under external injection~\cite{Shimizu1988, SanMiguel1995, Loke2023, Martin1997, Liu2020oil}. Therefore, VCSELs may be suitable building Ising computers, where the state of each Ising spin can be encoded into the lasing polarization~\cite{Loke2023, Yuan2025e, Yuan2026}. Ideally for this situation, VCSELs would exhibit negligible anisotropy and the lasing thresholds of the two polarizations would be identical. However, owing to strain in the quantum wells and the fabrication-induced geometric anisotropies, VCSELs are typically anisotropic in practice, presenting a non-neglectable dichroism~\cite{Shimizu1988, Mizutani1998}. This anisotropy causes VCSELs to preferentially lase in a certain polarization state, thereby constraining their potential applications in optical Ising computing. Although there have been reports of polarization locking through changing the shape of oxidation apertures~\cite{Yuan2025a, Yoshikawa1998, Yuan2025b, Zhang2018eye, Yuan2025c, Zhang2021asym}, rotating the orientation of the etched mesa~\cite{Yuan2025d}, or by integrating optical gratings~\cite{Ser1995, Huang2007}, additional approaches are desirable, which may lead to greater fine control of a VCSEL's polarization.

A metasurface is a type of 2D-like surface composed of micro- or nano- structures, which may provide the desired strong polarization control over traditional optical elements. In recent years, numerous works based on metasurfaces for polarization control have been reported. In addition, versatile applications have been realized by integrating metasurfaces and VCSELs, like generating programmable structured light~\cite{Wu2024}, modulating the output beam~\cite{Ni2022, Wen2023}, detecting the polarization state of the output light~\cite{Wen2021}, modulating the ratio of left-handed circularly polarized (LCP) and right-handed circularly polarized (RCP) components in the lasing light~\cite{Jia2023}, enhancing the quality factor with a reduced thickness of Bragg mirrors~\cite{Xu2020}, etc.

In this work, we proposed a metasurface-integrated VCSEL with lowered anisotropy, which can be utilized for all-optical Ising computing. We first introduce the design and working principles along with a detailed numerical analysis of the entire system. The metasurface itself is then fabricated on a GaAs substrate and its performance is verified. Although integration of the metasurface and VCSEL is left for future work in terms of fabrication, experimental testing of the fabricated metasurface gives us a high confidence that the system would function as desired. The idea is that by compensating for the reflectivity disparity between the two polarizations, i.e., bireflectance, the threshold gain difference is reduced, leading to a decrease in the anisotropy of the metasurface-integrated VCSEL. Our numerical simulations demonstrate that this anisotropy can be tuned by varying the geometric parameters of nanopillars within the metasurface. Near-field enhancement can be utilized to explain this bireflectance behavior. In the third section, we present the potential of our proposed metasurface-integrated VCSEL in the application of photonic Ising machines. By implementing mutual injection terms to the spin-flip model, we shall demonstrate that reducing VCSEL anisotropy can more than double the computational accuracy of a targeted photonic Ising machine. This work shows the promise that the combination of metasurfaces and VCSEL might someday bring to an application in all-optical Ising computing based on an injection-locking optical system.

\section{Design, fabrication and working principles}

The output light polarization state of a VCSEL is determined by the threshold gain, which refers to the gain value required to compensate for the total cavity losses when the laser reaches its threshold (i.e., begins to generate stimulated emission). If the threshold gain for a certain polarization is lower, the VCSEL will tend to lase polarized along that direction. Described as
\begin{equation*}
g_{th} = \frac{1}{\Gamma}\left[\alpha + \frac{1}{2L_{eff}} \ln\!\left(\frac{1}{R_1 R_2}\right)\right],
\end{equation*}
the threshold gain $g_{th}$ of a VCSEL is directly related to the reflectivity of the top mirror~\cite{Jia2023}. Here, $\Gamma$ is the confinement factor; $\alpha$ is the internal loss caused by the gain material and diffraction; $L_{eff}$ is the effective cavity length; and $R_1$ and $R_2$ are the reflectivities of the top mirrors and bottom mirrors, respectively. Therefore, any dichroism observed in VCSELs can be either enhanced or reduced by independently tuning the reflectivity difference between the two polarization states, i.e., the reflectivities of the X or Y polarization. By introducing a metasurface capable of doing this, the gain anisotropy of the integrated VCSELs can be effectively reduced, thereby decreasing the likelihood of lasing preferentially in one specific polarization state when increasing the current above threshold. However, it is important to note that achieving perfectly identical threshold gains is practically unfeasible, and VCSELs will still typically lase in either X or Y polarization---the goal, though, is to make the gains as close as possible so that small amounts of injection (in a hypothetical interconnected Ising system) could nudge the polarization one way or the other. In our simulations, the ``lasing polarization'' of a VCSEL is identified as the direction corresponding to the stronger polarization component rather than the vector sum.

\begin{figure}[!htbp]
\centering
\includegraphics[width=0.66\linewidth]{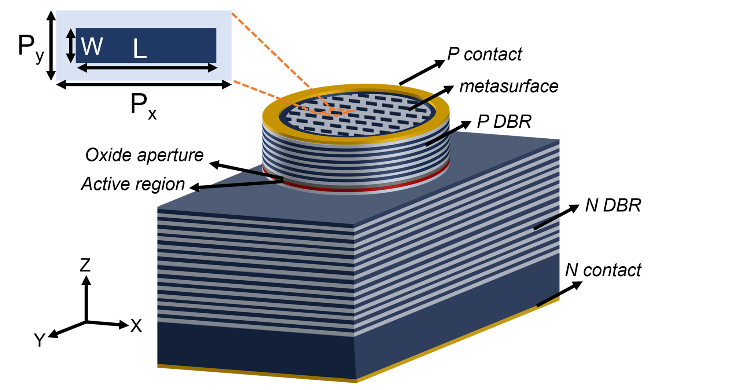}
\caption{Schematic diagram of the metasurface-integrated VCSEL, where the illustration in the upper-left shows the enlarged view of the metasurface within one period.}
\label{fig:1}
\end{figure}

Fig.~\ref{fig:1} shows a sketch of the GaAs-based metasurface-integrated VCSEL. This VCSEL is nominally to lase at 850~nm, and in the simplified model for simulating the effect of the metasurface, is comprised of 16 pairs of top distributed Bragg reflector (P-DBR) and 40 pairs of bottom DBR (N-DBR). Each DBR pair is composed of two AlGaAs layers with differing aluminum alloy contents, having refractive indices of 3.06 and 3.49, respectively~\cite{Gehrsitz2000}. The compensation for gain threshold of the VCSEL is based on the bireflectance induced by the metasurface. This metasurface consists of rectangular nano-pillars etched into the top GaAs layer, which is integrated into the top mirror stack and is designed to create an effective optical path of $\lambda/4$ thickness. The upper-left insert shows a top view of the metasurface (one period) with a rectangular nano-pillar unit. Here, we define a scaling factor $S$, and for metasurfaces of different geometric structures we define dimensions $P_y = S\times 200$~nm, $W = S\times 100$~nm, where $P_x$ and $L$ are kept as 1000~nm and 800~nm, respectively.

In our numerical model, the DBR layers are simplified as uniaxial to model anisotropy, exhibiting slight refractive index differences along the X and Y directions. Specifically, the layer with higher aluminum content has refractive indices of 3.06 and 3.08 along the X and Y axes, respectively, while the layer with lower aluminum content exhibits indices of 3.49 and 3.53 along the corresponding directions.

\begin{figure}[!htbp]
\centering
\includegraphics[width=0.92\linewidth]{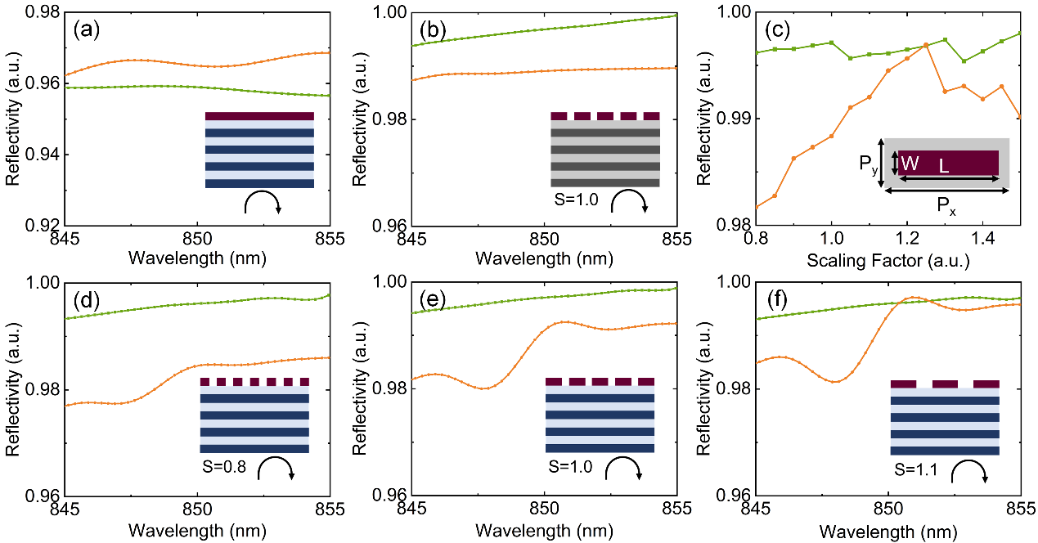}
\caption{Performance of the metasurface-integrated top DBR layers; the green and orange curves represent the reflectivities under X- and Y-polarized incidence, respectively. The light blue and dark blue layers represent anisotropic DBR layers with different aluminum alloy contents, while the light grey and the dark grey layers represent anisotropic DBR layers with different aluminum contents. The top GaAs layer is shown in purplish red. (a)(b) Bireflectance spectra of the top mirrors composed of anisotropic DBR layers and metasurface-integrated isotropic DBR layers, respectively. (c) Bireflectance of isotropic DBR layers at 850~nm when integrated with metasurface with different scaling factors. (d)--(f) Bireflectance spectra of metasurface-integrated anisotropic DBR layers around 850~nm. From left to right, the value of scaling factor $S$ is 0.8, 1.0, and 1.1, respectively.}
\label{fig:2}
\end{figure}

Reflectivity spectra of a set of top DBR mirrors were calculated via finite-difference-time-domain (FDTD) simulations around 850~nm. The metasurface is considered periodical with infinite duty cycles along both the X and Y axes. As shown in Fig.~\ref{fig:2}, the green and orange curves represent reflectivities of the top DBR mirrors under X- and Y-polarized light incidence, referred as $R_x$ and $R_y$, respectively. In Fig.~\ref{fig:2}(a), $R_y$ is higher than $R_x$. This disparity would result in a lower threshold gain for Y-polarized light, leading a VCSEL to preferentially lase in the Y-polarization state. This reflectivity dichroism can be compensated by a metasurface with a higher reflectivity for X-polarized light. Fig.~\ref{fig:2}(b) presents the bireflectance spectra of the metasurface-integrated isotropic DBR layers when the scaling factor $S$ is 1. In this case, the X-polarized light has a higher reflectivity than Y-polarized light. Furthermore, as illustrated in Fig.~\ref{fig:2}(c), at the lasing wavelength of 850~nm, the bireflectance induced by the metasurface can be tuned by adjusting the scaling factor of the rectangular nano-pillars. This modulation depends on the geometric configuration, thereby enabling precise control over the output polarization of this metasurface-integrated VCSEL. This bireflectance compensation induced by the metasurface reaches its maximum value when $S$ is 0.8. And generally, this bireflectance effect gradually decreases as $S$ increases to 1.25, and then continues to rise as $S$ increases further.

Figs.~\ref{fig:2}(d)--(e) are the reflectivity spectra for the metasurface-integrated anisotropic DBR, with $S = 0.8$, 1.0 and 1.1, respectively. In this case, we ignored the anisotropy of the top GaAs layer and regarded the index of the metasurface to be isotropic. When $S = 0.8$, the metasurface overcompensates, resulting in a preference for X-polarized lasing of the integrated VCSEL. This tendency diminishes progressively as $S$ increases to 1.0 and 1.1, consistent with the bireflectance modulation result shown in Fig.~\ref{fig:2}(c).

To quantify the bireflectance of the top mirrors, we define the dichroism factor $D = (R_x - R_y)/(R_x + R_y)$. Without metasurface integration, $D$ is $-0.31\%$ for the anisotropic DBR layers, indicating a preference for Y-polarized lasing. When the top DBR is integrated with the metasurface, the dichroism factor turns to 0.58\%, 0.28\%, and 0.06\% for $S = 0.8$, 1.0, and 1.1, respectively.

\begin{figure}[!htbp]
\centering
\includegraphics[width=0.66\linewidth]{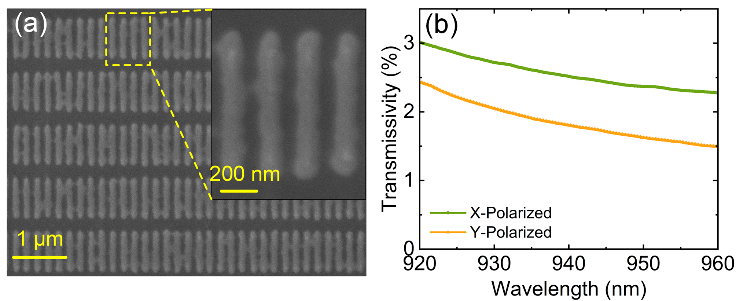}
\caption{(a) SEM images of the top view of the fabricated metasurface when $S = 1$. The embedded illustration shows a magnified view of four periods. (b) The experimental transmission spectra of the metasurface in the 900--1000~nm wavelength range.}
\label{fig:3}
\end{figure}

We fabricated a metasurface ($S = 1$) on a GaAs substrate to experimentally demonstrate its anisotropic reflectivity over two polarizations. The scanning electron microscopy (SEM) image of the metasurface shown in Fig.~\ref{fig:3}(a) indicates that the overall sample structures are similar to those designed, and the embedded SEM image depicts a partial magnification of four periods. A detailed fabrication process is presented in the Supplementary Material (Section~\ref{sec:S-fab}).

This GaAs-based metasurface was characterized using a UV-Vis-NIR Spectrophotometer (SolidSpec-3700, Shimadzu). This spectrophotometer only supports measurements in transmission mode. To avoid difficulties associated with measuring transmission near the GaAs band edge, we measured transmission at a slightly red-shifted 900--1000~nm wavelength range. As shown in Fig.~\ref{fig:3}(b), in this waveband, the transmissivity for X-polarized light (green curve) is higher than that of Y-polarized light (orange curve) while the reflectivity shows the opposite trend, which is in accordance with the simulation results. The low transmissivity is an artifact of the measurement; the beam size exceeded the fabricated area, so only the relative transmissivities can be compared.

\begin{figure}[!htbp]
\centering
\includegraphics[width=0.62\linewidth]{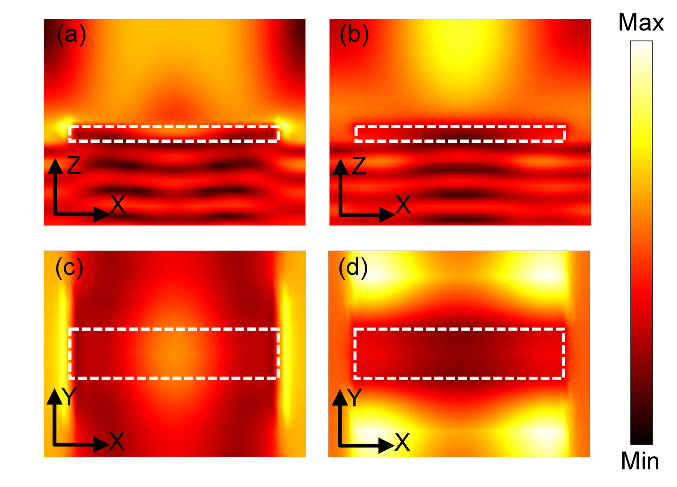}
\caption{(a)(c) Electric field distributions of the metasurface under X-polarized incidence in XZ and XY planes, respectively. (b)(d) Electric field distributions of the metasurface under Y-polarized incidence in XZ and XY planes, respectively. The dashed white lines indicate the position of the nanopillar.}
\label{fig:4}
\end{figure}

Near-field electric field distributions were analyzed in simulation to investigate the origin of the bireflectance caused by the metasurface at 850~nm. Here, the DBR pairs are regarded as isotropic and $S = 1$. Figs.~\ref{fig:4}(a) and (c) show the electric field distributions of the metasurface (a nanopillar unit is outlined by white dashed line) under X-polarized incidence along the XZ and XY planes, and Figs.~\ref{fig:4}(b) and (d) present those under Y-polarized incidence, respectively. For X-polarized light, the energy is predominantly concentrated along the short edges of the nanopillars. In contrast, the coupling effect is enhanced between adjacent nanopillars with a Y-polarized incidence. In this case, more energy is confined near the nanopillars and less energy is reflected.

\section{Metasurface-integrated VCSEL for all-optical Ising computing}

In fact, aside from manipulating the gain threshold, the birefringence of VCSELs can also be tuned through metasurface integration (Fig.~\ref{fig:S1}, Supplementary Material). Therefore, the polarization anisotropy of the VCSELs can be effective controlled using our design. In this section, we present the potential of metasurface-integrated VCSELs for all-optical Ising computing through simulation. This is shown based on injection-locking theory and the spin-flip model, which demonstrates that the polarization of a laser can flip into that of externally-injected light of a very close frequency~\cite{SanMiguel1995, Martin1997, Liu2020oil}. Under external and mutual injection, where injection between two lasers $i$ and $j$ corresponds to the interaction term $J_{i,j}$ in Equation~\eqref{eq:ising}, a system of VCSELs will naturally evolve toward the ground state through mode competition.

VCSELs can be utilized as Ising spins if they satisfy these characteristics:
\begin{enumerate}
\renewcommand{\labelenumi}{\alph{enumi})}
\item The VCSELs are easy to flip to the orthogonal polarization and are easy to flip back to their initial polarization under external injection.
\item All the VCSELs present the same behavior under the same external injection.
\end{enumerate}

As shown above, the polarization anisotropy of VCSELs can be controlled through metasurface integration. Here, to illustrate the potential of this scheme, we now analyze the time response of slave lasers with varying anisotropy under external injection of a constant amplitude based on the rate equations (Supplementary Material, Eqs.~\eqref{eq:S1}--\eqref{eq:S6}). Parameters used in our calculations are taken from Ref.~\cite{AlSeyab2011}. Specifically, the anisotropy of VCSELs is characterized by cavity dichroism ($\gamma_a$) and cavity birefringence ($\gamma_p$). Two VCSEL types shall be examined. In the isotropic case, both $\gamma_a$ and $\gamma_p$ are set to zero, while in the anisotropic case, $\gamma_a$ takes values of 0.67~ns$^{-1}$ and $-0.67$~ns$^{-1}$, and $\gamma_p$ takes values of 192~ns$^{-1}$ and $-192$~ns$^{-1}$, respectively.

\begin{figure}[!htbp]
\centering
\includegraphics[width=0.62\linewidth]{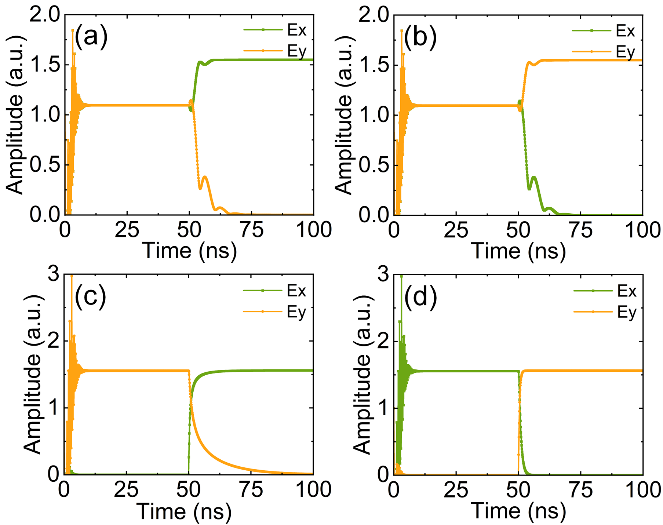}
\caption{The time evolution of lasing polarization of VCSELs under external injection. The time response for isotropic VCSEL under X- (a) and Y- polarized injection (b), respectively. The time response for anisotropic VCSELs under X- (c) and Y- polarized injection (d), respectively. In Figs.~\ref{fig:5}(c) and (d), $\gamma_a$ are set to 0.67~ns$^{-1}$ and $-0.67$~ns$^{-1}$, and $\gamma_p$ are set to 192~ns$^{-1}$ and $-192$~ns$^{-1}$, respectively.}
\label{fig:5}
\end{figure}

We first analyze the time response of one VCSEL under external injection: Initially, the amplitudes of the X- and Y-polarized components for each free-running isotropic VCSEL are identical. Upon external injection (starting at 50~ns, injection ratio $E_{inj}/E_{o} = 0.01$), the polarization of the VCSEL is locked to that of the external injection, as shown in Figs.~\ref{fig:5}(a) and (b). In contrast, Figs.~\ref{fig:5}(c) and (d) depict the polarization flipping of an anisotropic VCSEL when $E_{inj}/E_{o} = 0.1$. $\gamma_a$ is set to 0.67~ns$^{-1}$ and $-0.67$~ns$^{-1}$ in Figs.~\ref{fig:5}(c) and \ref{fig:5}(d), respectively; and $\gamma_p$ is set to 192~ns$^{-1}$ and $-192$~ns$^{-1}$, respectively. Initially, with positive values of $\gamma_a$ and $\gamma_p$, the free-running anisotropic VCSEL predominantly emits in Y polarization while negative values of $\gamma_a$ and $\gamma_p$ leads to lasing in X polarization. Here, $E_{inj}$ is the amplitude of the master laser. And $E_{o}$ denotes either $E_x$ or $E_y$, are the amplitude of X- or Y-polarized light of the slave laser at its stable state. The difference in angular frequency between that of the slave laser and the master laser is set to zero.

The isotropic VCSEL exhibits identical response to injection of either polarization, and its polarization will align with that of the master laser when $E_{inj}/E_{o}$ exceeds 0.0025. However, for an anisotropic VCSEL, this value must be increased to 0.07 to achieve polarization flipping---28 times higher than the threshold for the isotropic VCSEL, indicating that VCSELs with lowered anisotropy, as achieved via metasurface integration, are much easier to be locked. Further evidence comes from the time response of slave lasers with varying anisotropy under periodical injection (Fig.~\ref{fig:S2}, Supplementary Material). Again, despite the fact that the anisotropy of VCSELs can be greatly lowered down via metasurface integration, real VCSELs are never perfectly isotropic and zero values of $\gamma_a$ and $\gamma_p$ are only used for demonstration purposes.

\begin{figure}[!htbp]
\centering
\includegraphics[width=0.66\linewidth]{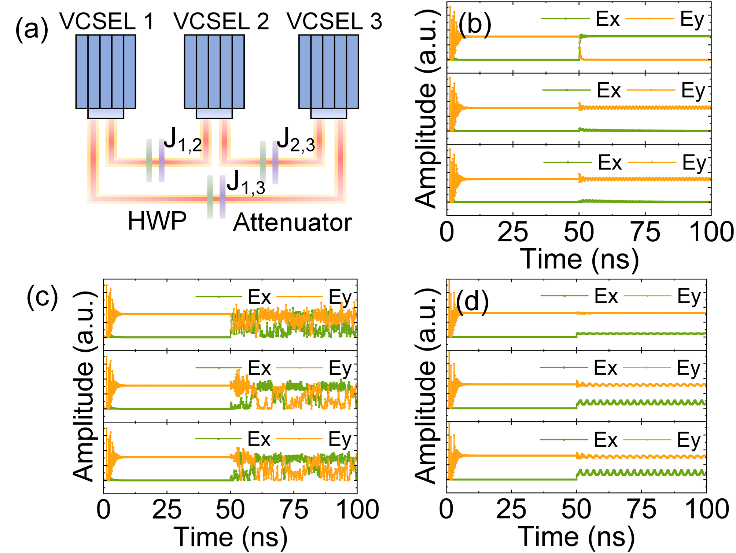}
\caption{Impact of VCSELs' anisotropy on the computational accuracy in 3-bit photonic Ising machines. (a) Schematic diagram of the optical setup. (b)--(d) The evolution of the 3-bit Ising system composed of VCSELs with varying anisotropy under a random interaction matrix $J$.}
\label{fig:6}
\end{figure}

Next, we utilize a metasurface-integrated VCSELs system in a 3-bit Ising machine to demonstrate its potential in all-optical computing, where each VCSEL's polarization state corresponds to an ``Ising spin''. The optical setup for the 3-bit photonic Ising computer is illustrated in Fig.~\ref{fig:6}(a), inspired by Ref.~\cite{Utsunomiya2011}. The Hamiltonian value is calculated as in Eq.~\eqref{eq:ising}. $\sigma$ represents the polarization state of each bit, and is defined as $+1$ for X polarization and $-1$ for Y polarization. The Zeeman term $h_i$, which accounts for the influence of the environment on bit $i$, is set to zero in our calculations.

The computational results of the photonic Ising machine are uniquely determined by the interaction matrix $J$, with interaction strength and polarity hypothetically controlled via attenuators and half-wave plates (HWP) between VCSELs. Specifically, $J_{i,j}$ defines the coupling between VCSEL $i$ and VCSEL $j$. A positive $J_{i,j}$ couples the X-polarized component of VCSEL $i$ to VCSEL $j$'s X polarization, while a negative $J_{i,j}$ couples it to VCSEL $j$'s Y polarization, and vice versa. Consequently, two directly linked VCSELs lase in the same direction if $J_{i,j} > 0$ and in opposite polarizations for $J_{i,j} < 0$.

According to our design, to guarantee uniform response across all Ising bits under injection, the parameters of each VCSEL are kept identical. The time response for each bit can be described by implementing mutual injection terms into the laser rate equation, which is described in the Supplementary Material. A detailed derivation can be found in our group's previous work~\cite{Loke2023}. To simplify the process, the coupling coefficient of the X polarization from spin $i$ to spin $j$, (denoted as $K_{inj,x_{i,j}}$ and $K_{inj,y_{i,j}}$, see Supplementary Material, Eqs.~\eqref{eq:S7}--\eqref{eq:S16}), are uniformly set to 3.55~ns$^{-1}$ for all combinations of spin $i$ and spin $j$; the frequency detuning between all the VCSELs is considered zero in our study.

The anisotropy of VCSELs can be tuned by integrating a metasurface with different scaling factors. To assess its influence on Ising computation accuracy, we compared the performance of photonic Ising machines composed of VCSELs with varying anisotropy. Here, a symmetric matrix $J$ was constructed with zero diagonal elements and off-diagonal elements uniformly distributed in the interval $[-1, 1]$, configured as
\begin{equation*}
J = \begin{bmatrix} 0 & 0.86 & 0.73 \\ 0.86 & 0 & -0.95 \\ 0.73 & -0.95 & 0 \end{bmatrix}.
\end{equation*}
The correct solutions for this Ising problem are $[+1, -1, -1]$ and $[-1, +1, +1]$.

Figs.~\ref{fig:6}(b)--(d) depict the time evolution of the system under mutual injection beginning at 50~ns. From left to right $\gamma_a$ of VCSELs in the systems are set to 0.67~ns$^{-1}$, 0.67~ns$^{-1}$, and 6.7~ns$^{-1}$, respectively; $\gamma_p$ are set to 192~ns$^{-1}$, 500~ns$^{-1}$, and 1920~ns$^{-1}$, respectively. When the VCSELs have the lowest anisotropy, the system can easily evolve into the correct solution. As the anisotropy of VCSELs increases, fluctuations are observed but the system can still converge. However, when the anisotropy of the VCSELs is large enough, they tend to lase along their preferred polarization orientation. In this case, their capability to undergo polarization flipping under external injection diminishes, leading to a decline in calculation accuracy. For reproducibility verification, repetitive tests are carried out under 100 different interaction matrices, with all matrices constrained to be symmetric with zero diagonal elements and all off-diagonal elements bounded between $-1$ and $+1$. The calculation accuracy decreased from 62\% to 30\% as the anisotropy increases, demonstrating the improved performance of VCSELs in optical computation after metasurface integration. Since the output power of each VCSEL has to be split to multi-channels to complete optical computing, the decrease in required injection power enabled by metasurface integration is of great importance for solving large-scale Ising problems. Additionally, in cases where the Ising problem has multiple solutions, the initial anisotropy of the VCSELs will drive the system towards a certain solution (Fig.~\ref{fig:S3}, Supplementary Material).

\section{Conclusion}

In summary, we proposed a metasurface-integrated VCSEL, where the metasurface is incorporated into the top mirrors to compensate for the bireflectance induced by the VCSEL's anisotropy. The polarization of the VCSEL can be directly controlled by the bireflectance value of the top mirrors, which is adjustable by tuning the geometric structures of nano-pillars. The bireflectance of the metasurface is evidenced through experiments and further explained via the near-field enhancement analysis. We then simulated an injection-locked system and demonstrated that VCSELs with lowered anisotropy are more easily and more stable to be locked. Our proposed metasurface-integrated VCSEL offers a promising paradigm for all-optical computation with Ising machines. We have demonstrated a 3-bit photonic Ising machine as an example. Because the metasurface is patterned directly into the top GaAs layer, the approach could in future be combined with array-level VCSEL fabrication~\cite{Yuan2025c, Yuan2025e} and co-integrated with other on-chip photonic components~\cite{Lim2025}.

\paragraph*{Funding.}
This work was supported by the National Research Foundation, Singapore, under its Competitive Research Programme (NRF CRP24-2020-0003) and by both the National Research Foundation, Singapore, and A*STAR under the Quantum Engineering Programme (NRF 2021-QEP2-02-P12). This work was also supported by National Natural Science Foundation of China (No.~52075517).

\paragraph*{Data availability.}
Data underlying the results presented in this paper are not publicly available at this time but may be obtained from the authors upon reasonable request.

\renewcommand{\refname}{References}
{\small
\bibliographystyle{ieeetr}
\bibliography{references}
}

\clearpage
\setcounter{section}{0}
\setcounter{equation}{0}
\setcounter{figure}{0}
\renewcommand{\thesection}{S\arabic{section}}
\renewcommand{\theequation}{S\arabic{equation}}
\renewcommand{\thefigure}{S\arabic{figure}}

\begin{center}
{\Large\bfseries Supplementary Material}\\[4pt]
{\large Metasurface-integrated VCSEL designed for polarization control in optical Ising machines}
\end{center}

\section{Metasurface fabrication}
\label{sec:S-fab}

The fabrication of metasurface-integrated VCSEL would consist of two parts---the fabrication of the VCSEL and the fabrication of the metasurface on the top of the VCSEL. In this work, we only fabricated a metasurface on a GaAs substrate to demonstrate its capabilities but fabrication on a VCSEL facet would be similar.

First, electron beam lithography (EBL) resist (ma-N 2401) was spin-coated at speed of 6000~rpm for 45 seconds. A pre-bake was then performed at 90\,\textdegree C for 1 minute to enhance adhesion and remove impurities and bubbles. EBL was carried out at an exposure step size of 0.004~$\mu$m with an exposure dose of 120~$\mu$C/cm$^2$. The fabrication area is 50~$\mu$m $\times$ 50~$\mu$m. After exposure, a post-bake at 90\,\textdegree C for 1 minute was applied to stabilize the resist pattern. The sample was then developed in ma-D525 solution for 2 minutes and 50 seconds, and rinsed in deionized water for 5 minutes.

The ma-N 2401 is a negative resist. The unexposed areas were removed after development while the exposed regions remained. To transfer the pattern from the resist to the GaAs substrate, we conducted etching using inductively coupled plasma (ICP). Cl$_2$ and BCl$_3$ were utilized as the etching gases and the flow ratio was set to 1:20 (in sccm). The etching time is 60 seconds, with an etching power of 500~W and a gas pressure of 10~mTorr. Subsequently, a plasma asher was employed for resist removal. Oxygen was used as the reactive gas, with an etching time of 15 minutes, a gas flow rate of 50~sccm, and a pressure of 5~Pa.

The fabricated rectangular nano-pillars were 770~nm in length and 140~nm in width, with periods of 1000~nm and 210~nm along the X and Y directions (output emission along Z direction), respectively. The height was 61.5~nm, demonstrating good agreement with the design values.

\section{Spin-flip model under external injection and mutual injection}

By implementing external injection components from the master laser, the rate function of the slave VCSELs can be described as follows:
\begin{align}
\frac{dE_x}{dt} &= \kappa\left[(N-1)E_x - mE_y\left(\sin\Delta\phi + \alpha\cos\Delta\phi\right)\right] - \gamma_a E_x + K_{inj,x}E_{inj,x}\cos\Delta x \label{eq:S1}\\
\frac{dE_y}{dt} &= \kappa\left[(N-1)E_y + mE_x\left(\alpha\cos\Delta\phi - \sin\Delta\phi\right)\right] + \gamma_a E_y + K_{inj,y}E_{inj,y}\cos\Delta y \label{eq:S2}\\
\frac{d\phi_x}{dt} &= \kappa\left[\alpha(N-1) + m\frac{E_y}{E_x}\left(\cos\Delta\phi - \alpha\sin\Delta\phi\right)\right] - \Delta\omega - \alpha\gamma_a + K_{inj,x}\frac{E_{inj,x}}{E_x}\sin\Delta x \label{eq:S3}\\
\frac{d\phi_y}{dt} &= \kappa\left[\alpha(N-1) - m\frac{E_x}{E_y}\left(\alpha\sin\Delta\phi + \cos\Delta\phi\right)\right] - \Delta\omega + \alpha\gamma_a + K_{inj,y}\frac{E_{inj,y}}{E_y}\sin\Delta y \label{eq:S4}\\
\frac{dN}{dt} &= -\gamma\left[N\left(1 + E_x^2 + E_y^2\right) - \eta - 2mE_yE_x\sin\Delta\phi\right] \label{eq:S5}\\
\frac{dm}{dt} &= -\gamma_s m - \gamma\left[m\left(E_x^2 + E_y^2\right)\right] + 2\gamma N E_y E_x\sin\Delta\phi \label{eq:S6}
\end{align}
Here, $E_x$ and $E_y$ are the amplitudes, and $\phi_x$ and $\phi_y$ are phases of the X- and Y- polarized components of the slave VCSEL, respectively. $N$ is the inversion carrier density; $m$ is the difference in carrier density between two magnetic sublevels; $\kappa$ is the photon decay rate, and $\gamma$ is the carrier decay rate; $\gamma_s$ is the spin relaxation rate; $\gamma_a$ is the cavity dichroism, i.e., the gain anisotropy representing for the difference in the gain-to-loss ratio between X- and Y-polarized modes; $\gamma_p$ is the cavity birefringence, which presents for the phase anisotropy and is responsible for the frequency splitting between the two modes; $\alpha$ is the linewidth enhancement factor; $\eta$ is the bias current normalized to the threshold current; $K_{inj,x}$ and $K_{inj,y}$ represents the injection coupling coefficients; $\Delta\omega$ is the difference in angular frequency between that of the slave laser and the master laser, and $\Delta\omega = \omega_{inj} - \omega_1$. $\omega_1$ is defined as the average angular frequency of X- and Y-polarized modes, and $\omega_x = \alpha\gamma_a - \gamma_p$ and $\omega_y = \gamma_p - \alpha\gamma_a$ represents for the angular frequency shift of the two modes from the injected signal, respectively. $\Delta x = \omega_y t - \phi_x$; $\Delta y = \omega_x t - \phi_y = (\alpha\gamma_a - \gamma_p)t - \phi_y$; $\Delta\phi = 2(\gamma_p - \alpha\gamma_a)t + \phi_y - \phi_x$. Therefore, the anisotropy of the VCSEL is characterized and determined by $\gamma_a$ and $\gamma_p$. Parameters used in our calculations are taken from Ref.~\cite{AlSeyab2011}, namely, $\gamma = 0.67$~ns$^{-1}$; $\gamma_s = 1000$~ns$^{-1}$; $\kappa = 125$~ns$^{-1}$; $\eta = 3.4$; $K_{inj,x}$ and $K_{inj,y} = 35.5$~ns$^{-1}$; $\alpha = 3$.

As shown in Fig.~\ref{fig:6}(a), the interaction polarity between VCSEL $i$ and VCSEL $j$ can be controlled by the presence or absence of a half wave plate. In this case, the $J$ matrix can be decomposed into $J = A - A_{HWP}$, where $A$ represents positive coupling and $A_{HWP}$ represents negative coupling. For $J_{i,j} > 0$, $A_{i,j} = J_{i,j}$ and $A_{HWP,i,j} = 0$. Otherwise, for $J_{i,j} < 0$, $A_{i,j} = 0$ and $A_{HWP,i,j} = -J_{i,j}$. Here, $0 \le A_{i,j} \le 1$, $0 \le A_{HWP,i,j} \le 1$. For instance, if $J$ matrix is given as
\begin{equation*}
J = \begin{bmatrix} 0 & 1 & -1 \\ 1 & 0 & 1 \\ -1 & 1 & 0 \end{bmatrix},
\end{equation*}
then $A$ and $A_{HWP}$ will take the form
\begin{equation*}
A = \begin{bmatrix} 0 & 1 & 0 \\ 1 & 0 & 1 \\ 0 & 1 & 0 \end{bmatrix}
\quad\text{and}\quad
A_{HWP} = \begin{bmatrix} 0 & 0 & 1 \\ 0 & 0 & 0 \\ 1 & 0 & 0 \end{bmatrix},
\end{equation*}
respectively. By implementing the mutual injection terms, the rate equation for each laser can be described as:
\begin{align}
\frac{dE_{x,i}}{dt} &= \kappa\left[(N_i-1)E_{x,i} - m_iE_{y,i}\left(\sin\Delta\phi_i + \alpha\cos\Delta\phi_i\right)\right] - \gamma_a E_{x,i} + K_{inj,x}E_{inj,x}\cos\Delta x + \sum_{j\neq i}^{M}\varepsilon_{x,ji} \label{eq:S7}\\
\frac{dE_{y,i}}{dt} &= \kappa\left[(N_i-1)E_{y,i} + m_iE_{x,i}\left(\alpha\cos\Delta\phi_i - \sin\Delta\phi_i\right)\right] + \gamma_a E_{y,i} + K_{inj,y}E_{inj,y}\cos\Delta y + \sum_{j\neq i}^{M}\varepsilon_{y,ji} \label{eq:S8}\\
\frac{d\phi_{x,i}}{dt} &= \kappa\left[\alpha(N_i-1) + m_i\frac{E_{y,i}}{E_{x,i}}\left(\cos\Delta\phi_i - \alpha\sin\Delta\phi_i\right)\right] - \alpha\gamma_a \nonumber\\
&\quad + K_{inj,x}\frac{E_{inj,x}}{E_x}\sin\Delta x + \sum_{j\neq i}^{M}\varnothing_{x,ji} \label{eq:S9}\\
\frac{d\phi_{y,i}}{dt} &= \kappa\left[\alpha(N_i-1) - m_i\frac{E_{x,i}}{E_{y,i}}\left(\alpha\sin\Delta\phi_i + \cos\Delta\phi_i\right)\right] + \alpha\gamma_a \nonumber\\
&\quad + K_{inj,y}\frac{E_{inj,y}}{E_y}\sin\Delta y + \sum_{j\neq i}^{M}\varnothing_{y,ji} \label{eq:S10}\\
\frac{dN_i}{dt} &= -\gamma\left[N_i\left(1 + E_{x,i}^2 + E_{y,i}^2\right) - \eta - 2m_iE_{y,i}E_{x,i}\sin\Delta\phi_i\right] \label{eq:S11}\\
\frac{dm_i}{dt} &= -\gamma_s m_i - \gamma\left[m_i\left(E_{x,i}^2 + E_{y,i}^2\right)\right] + 2\gamma N_i E_{y,i}E_{x,i}\sin\Delta\phi_i \label{eq:S12}
\end{align}
\begin{align}
\sum_{j\neq i}^{M}\varepsilon_{x,ji} &= \sum_{j}^{\text{no HWP}}\varepsilon_{xx,ji} + \sum_{j}^{\text{HWP}}\varepsilon_{yx,ji} \nonumber\\
&= \sum_{j}^{\text{no HWP}} K_{inj,x_{i,j}}\, a_{ji}\, E_{inj,x_j}\cos(\phi_{x,j} - \phi_{x,i}) \nonumber\\
&\quad + \sum_{j}^{\text{HWP}} K_{inj,x_{i,j}}\, a_{HWP,ji}\, E_{inj,x_j}\cos(w_y - w_x + \phi_{y,j} - \phi_{x,i}) \label{eq:S13}\\[4pt]
\sum_{j\neq i}^{M}\varepsilon_{y,ji} &= \sum_{j}^{\text{no HWP}}\varepsilon_{yy,ji} + \sum_{j}^{\text{HWP}}\varepsilon_{xy,ji} \nonumber\\
&= \sum_{j}^{\text{no HWP}} K_{inj,y_{i,j}}\, a_{ji}\, E_{inj,y_j}\cos(\phi_{y,j} - \phi_{y,i}) \nonumber\\
&\quad + \sum_{j}^{\text{HWP}} K_{inj,y_{i,j}}\, a_{HWP,ji}\, E_{inj,y_j}\cos(w_x - w_y + \phi_{x,j} - \phi_{y,i}) \label{eq:S14}\\[4pt]
\sum_{j\neq i}^{M}\varnothing_{x,ji} &= \sum_{j}^{\text{no HWP}}\varnothing_{xx,ji} + \sum_{j}^{\text{HWP}}\varnothing_{yx,ji} \nonumber\\
&= \frac{1}{E_{x,i}}\sum_{j}^{\text{no HWP}} K_{inj,x_{i,j}}\, a_{ji}\, E_{inj,x_j}\sin(\phi_{x,j} - \phi_{x,i}) \nonumber\\
&\quad + \frac{1}{E_{x,i}}\sum_{j}^{\text{HWP}} K_{inj,x_{i,j}}\, a_{HWP,ji}\, E_{inj,x_j}\sin(w_y - w_x + \phi_{y,j} - \phi_{x,i}) \label{eq:S15}\\[4pt]
\sum_{j\neq i}^{M}\varnothing_{y,ji} &= \sum_{j}^{\text{no HWP}}\varnothing_{yy,ji} + \sum_{j}^{\text{HWP}}\varnothing_{xy,ji} \nonumber\\
&= \frac{1}{E_{y,i}}\sum_{j}^{\text{no HWP}} K_{inj,y_{i,j}}\, a_{ji}\, E_{inj,y_j}\sin(\phi_{y,j} - \phi_{y,i}) \nonumber\\
&\quad + \frac{1}{E_{y,i}}\sum_{j}^{\text{HWP}} K_{inj,y_{i,j}}\, a_{HWP,ji}\, E_{inj,y_j}\sin(w_x - w_y + \phi_{x,j} - \phi_{y,i}) \label{eq:S16}
\end{align}

\section{The birefringence behavior of VCSELs before and after metasurface integration}

Aside from the regulation ability over the threshold gain, i.e., the bireflectance discussed in the main text, the integrated metasurface also enables control over the cavity birefringence. Due to its anisotropic structure, the metasurface introduces different effective refractive indices along the X and Y directions. This allows the intrinsic birefringence induced by the anisotropic DBRs to be either enhanced or compensated, effectively tuning the birefringence parameter $\gamma_p$~\cite{Martin1997}.

\begin{figure}[!htbp]
\centering
\includegraphics[width=\linewidth]{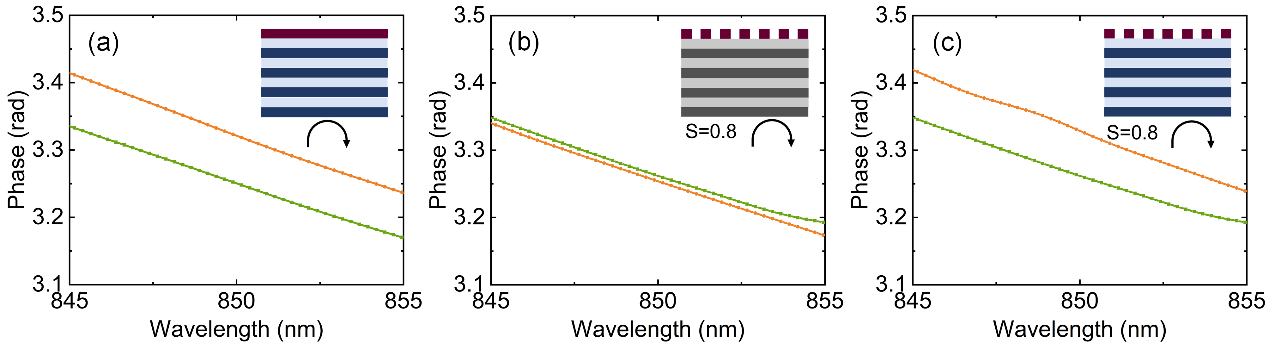}
\caption{The birefringence behavior of the metasurface-integrated top DBR layers. (a)(b) Birefringence spectra of the top mirrors composed of anisotropic DBR layers and metasurface-integrated isotropic DBR layers, respectively. (c) Birefringence spectra of metasurface-integrated anisotropic DBR layers. Here, the green and orange curves correspond to the phase under X- and Y-polarized incidence, respectively; the scaling factor of the metasurface is 0.8; the color coding of the layers are in accordance with those in Fig.~\ref{fig:2}.}
\label{fig:S1}
\end{figure}

Fig.~\ref{fig:S1}(a) depicts the birefringence spectra of the anisotropic DBR layers, calculated using the same parameters as in Fig.~\ref{fig:2}(a). As seen, the Y-polarized light accumulates a larger phase shift owing to the intrinsic birefringence of the VCSEL. This effect can be compensated by integrating a metasurface that introduces an opposite birefringence response. As shown in Figs.~\ref{fig:S1}(a) and (b), the birefringence of the top mirrors is reduced from 0.01078 to 0.01023 after metasurface integration---decreased by 5.1\%. Here, the scaling factor $S$ is set to 0.8 as an example.

\section{The lasing characteristics of anisotropic VCSELs under periodical injection}

\begin{figure}[!htbp]
\centering
\includegraphics[width=\linewidth]{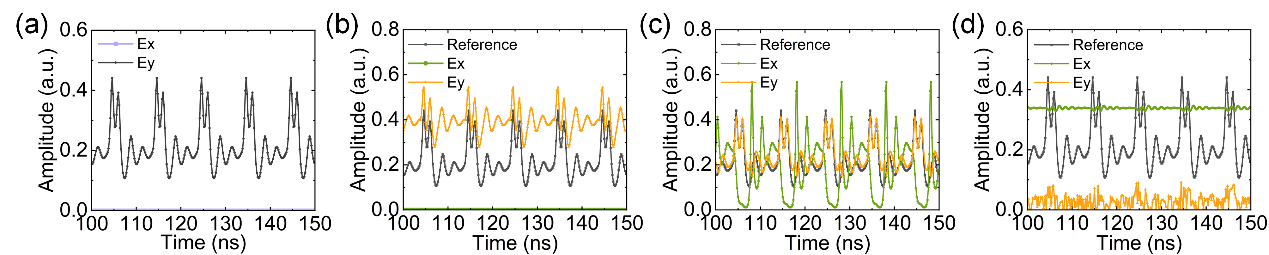}
\caption{The lasing characteristics of anisotropic VCSELs under periodical injections. (a) The time response of the master laser under Ultra-Gaussian bias current. (b)--(d) The time response of slave lasers with varying degrees of anisotropy under periodical injection from the master laser shown in (a). From left to right the anisotropy gradually increases.}
\label{fig:S2}
\end{figure}

Here, we investigated the time response of anisotropic slave lasers under periodic injection. Here, the bias current $\eta$ of the master laser follows an Ultra-Gaussian profile,
\begin{equation*}
\eta = \eta_0\left(1 + W e^{-\left(\frac{8(t - \tau/2)}{\tau}\right)^{2m}}\right),
\end{equation*}
where $\eta_0$ is the normalized DC bias current; $W$ is the modulation amplitude; $m$ is the order; and $\tau$ is the period; $n\tau \le t \le (n+1)\tau$ and $n$ is an integer. Parameter values are taken from Ref.~\cite{Yang2009}. Specifically, $\eta_0$ is 1.03; $W$ is 0.08; $m$ is 3; and $\tau$ is 10~ns. Under this bias current, the emitted light of the master laser is Y-polarized with a period of 10~ns, as shown in Fig.~\ref{fig:S2}(a). Figs.~\ref{fig:S2}(b)--(d) are the time evolution of the slave lasers with different anisotropy. The bias current $\eta$ of the slave lasers is set to 1.11 and the angular frequency difference $\Delta\omega$ is set to zero, with $\gamma_a$ fixed at $-0.67$~ns$^{-1}$. From left to right, $\gamma_p$ gradually decreases from $-192$~ns$^{-1}$, to $-215$~ns$^{-1}$, and $-500$~ns$^{-1}$. As described earlier, these anisotropic VCSELs initially lase along the X polarization. The magnitude of $\gamma_p$ reflects the ability of the slave VCSEL to maintain its initial polarization when subjected to an external injection of the orthogonal polarization.

When the absolute value of $\gamma_p$ is relatively small, the slave laser is easy to be locked by the master laser. In this case, the slave laser emits in Y polarization while the X-polarized component is completely suppressed, with a lasing period matching that of the master laser. However, as the absolute value of $\gamma_p$ increases to 215~ns$^{-1}$, the polarization state of the slave laser periodically switches between X and Y. With a strong external injection, the slave laser will lase in Y polarization, while a weaker injection leads to an X-polarized output. As the absolute value of $\gamma_p$ further increases to 500~ns$^{-1}$, it becomes significantly harder for the slave laser to be locked with the same injection intensity. Although slight fluctuations in the Y-polarized component are observed, the slave laser predominantly lases in X polarization with almost the same amplitude. Additionally, the enhanced thermal effect induced by higher injection power causes a redshift in the lasing wavelength of the slave laser. As a result, the wavelength mismatch between the master laser and slave laser increases. If a temperature controller is used to maintain the master laser at its initial wavelength, a higher injection power will further be required to achieve injection locking of the slave laser due to the increased detuning. Therefore, in accordance with the behavior under VCSELs with lower anisotropy are easier to be locked, and are better suited for applications such as all-optical Ising computation.

\section{The influence of the anisotropy of VCSEL on the evolutionary preference in the presence of multiple solutions}

Finally, we investigate how the initial polarization state of free-running VCSELs, influenced by their anisotropy, affects the computational preference when the Ising problem has multiple solutions. In this case, we compared the calculation results of the 3-bit Ising system consisting VCSELs with $\gamma_a = 0.67$~ns$^{-1}$ and $\gamma_p = 192$~ns$^{-1}$, and $\gamma_a = -0.67$~ns$^{-1}$ and $\gamma_p = -192$~ns$^{-1}$. The free-running VCSELs lase along Y- and X- polarization, respectively.

\begin{figure}[!htbp]
\centering
\includegraphics[width=0.82\linewidth]{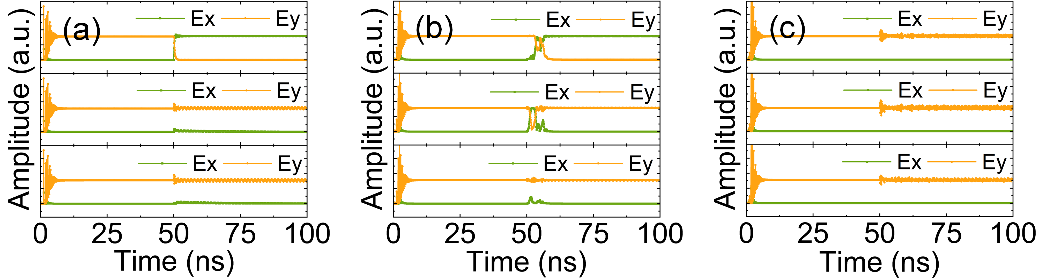}
\caption{The evolution of the 3-bit Ising system under three random interaction matrices. Here, $\gamma_a$ and $\gamma_p$ of the VCSELs are set to 0.67~ns$^{-1}$ and 192~ns$^{-1}$, respectively.}
\label{fig:S3}
\end{figure}

To better demonstrate the impact of the anisotropy of VCSELs on the evolutionary preference, we randomly selected and configured three $J$ matrices as
\begin{equation*}
J_1 = \begin{bmatrix} 0 & 0.86 & 0.73 \\ 0.86 & 0 & -0.95 \\ 0.73 & -0.95 & 0 \end{bmatrix},\quad
J_2 = \begin{bmatrix} 0 & 0.9 & 0.25 \\ 0.9 & 0 & -0.3 \\ 0.25 & -0.3 & 0 \end{bmatrix},\quad
J_3 = \begin{bmatrix} 0 & -0.92 & -0.02 \\ -0.92 & 0 & -0.66 \\ -0.02 & -0.66 & 0 \end{bmatrix},
\end{equation*}
respectively. Correspondingly, the correct solutions are $[+1, -1, -1]$ or $[-1, +1, +1]$ for $J_1$, $[+1, -1, -1]$ or $[-1, +1, +1]$ for $J_2$, and $[-1, -1, -1]$ or $[+1, +1, +1]$ for $J_3$, respectively. Figs.~\ref{fig:S3}(a)--(c) illustrate the time response of the 3-bit Ising system consisting VCSELs with $\gamma_a = 0.67$~ns$^{-1}$ and $\gamma_p = 192$~ns$^{-1}$ under $J_1$, $J_2$, $J_3$, respectively. In all three cases, the optical system evolves to the correct solution.

\begin{figure}[!htbp]
\centering
\includegraphics[width=\linewidth]{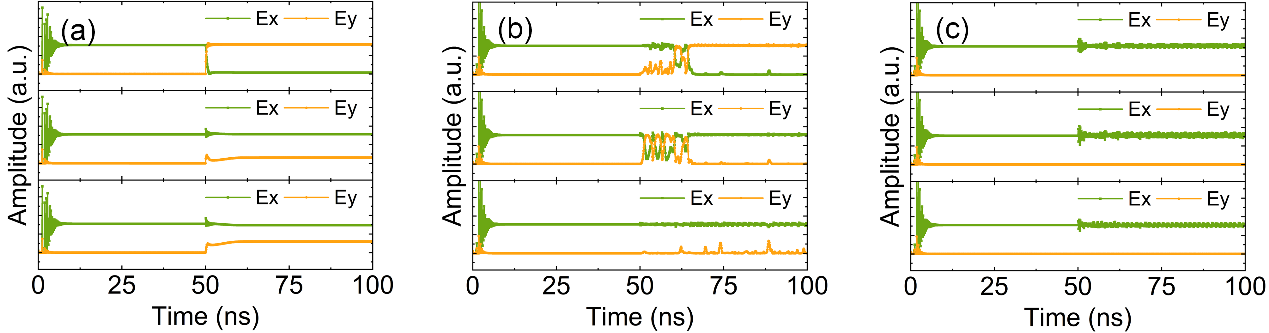}
\caption{The influence of the anisotropy of VCSEL on the evolutionary preference in the presence of multiple solutions. Here, the interaction matrices are kept as same as ones shown in Figs.~\ref{fig:S3}(a)--(c). $\gamma_a$ and $\gamma_p$ are replaced by their opposite values in Fig.~\ref{fig:S3}, while other parameters remain unchanged.}
\label{fig:S4}
\end{figure}

As shown in Fig.~\ref{fig:S4}(a)--(c), the optical system can still converge to a correct solution after mutual injection if the 3-bit Ising system is composed of VCSELs with $\gamma_a = -0.67$~ns$^{-1}$ and $\gamma_p = -192$~ns$^{-1}$, but the evolutionary results differ from those shown in Figs.~\ref{fig:S3}. For instance, this optical system initially resembles the configuration of $[+1, +1, +1]$. Although both $[-1, -1, -1]$ and $[+1, +1, +1]$ are valid solutions for $J_3$, the system is less likely to evolve into $[-1, -1, -1]$ due to the influence of VCSELs' anisotropy. This suggests that the anisotropy biases the system toward a specific solution when multiple solutions are possible. For photonic Ising machines composed of VCSELs lasing in X and Y polarization, two different solutions may be obtained.

\end{document}